\UseRawInputEncoding
\documentclass[aps,prb,twocolumn,superscriptaddress]{revtex4-1}
\usepackage{graphicx,bm,amssymb,amsmath,xcolor,pifont,soul}
\usepackage[linktocpage=true,colorlinks=true,pdfborder={0 0 0},linkcolor=blue,citecolor=red,filecolor=yellow,urlcolor=blue,bookmarks,pdfauthor={},]{hyperref}
\usepackage{ulem}
\usepackage{multirow}

\begin{document}
\title{Stability, electronic quantum states, and magnetic interactions of Er$^{3+}$ ions in Ga$_2$O$_3$} 
\author{Yogendra Limbu}
\affiliation{Department of Physics and Astronomy, University of Iowa, Iowa City, Iowa 52242, USA}
\author{Hari Paudyal}
\affiliation{Department of Physics and Astronomy, University of Iowa, Iowa City, Iowa 52242, USA}
\author{Michael E. Flatt\'e}
\affiliation{Department of Physics and Astronomy, University of Iowa, Iowa City, Iowa 52242, USA}
\affiliation{Department of Applied Physics, Eindhoven University of Technology, Eindhoven, The Netherlands}
\author{Durga Paudyal}
\email{durga.quantum@gmail.com}
\affiliation{Department of Physics and Astronomy, University of Iowa, Iowa City, Iowa 52242, USA}
\begin{abstract}
\noindent We report an \textit{ab initio} study of phase stability, defect formation, electronic structure, and multiple magnetic, Dzyaloshinskii-Moriya, optical, hyperfine, and crystal field interactions in erbium (Er) doped wide band gap $\alpha$- and $\beta$-gallium oxides (Ga$_2$O$_3$), applicable to optoelectronic and quantum devices.
The chemical, structural, mechanical, and dynamical stabilities of the pristine phases are confirmed from respective negative formation energies, negative cohesive energies, favorable elastic constants, and positive phonon frequencies. The phonon dispersions indicate that the Ga-O bonds are uniform in the $\alpha$-phase, while they vary in the $\beta$-phase due to the anisotropic polyhedral movement. 
The defect formation energy analysis confirms that both Er-doped $\alpha$- and $\beta$-Ga$_2$O$_3$ prefer Er$^{3+}$ (neutral) state. The underestimated band gaps of the pristine phases from standard density functional theory (DFT) calculations as compared to experimental values are corrected by employing the hybrid functional calculations, resulting to indirect band gaps of 5.21~eV in $\alpha$-Ga$_2$O$_3$  and 4.94~eV in $\beta$-Ga$_2$O$_3$. The site preference energy analysis indicates partial occupation of Er in the octahedral site of Ga. Anisotropic nature of hyperfine tensor coefficients of Er are similar in both phases which may be due to the occupation of Er in the same octahedral Ga site. On the other-hand, calculated magnetic exchange interaction between two Er dopants is negative for $\alpha$ and positive for $\beta$, indicating antiferromagnetic ground state in the former and the ferromagnetic ground state in the latter. A large values of Dzyaloshinskii-Moriya interactions (DMIs) are obtained along the $x$ direction in the $\alpha$ and along the $y$ direction in the $\beta$.
The analysis of dielectric constants and refractive indices of both pristine and Er doped phases shows a good agreement with available experimental values. The calculated  optical anisotropy is slightly higher in $\beta$ than those in $\alpha$, which is due to the involvement of lower symmetry in $\beta$. The crystal field coefficients (CFCs) calculated from DFT are used to analyze 4$f$ multiplets and 4$f$ - 4$f$ transitions. Thus calculated lowest energy level of the first excited state to the lowest energy level of the ground state is about 1.53~$\mu$m, which is in a good agreement with available experiment and it falls within the quantum telecommunication wavelength range. 
\end{abstract}
\maketitle
\section{Introduction}
\noindent Gallium oxide (Ga$_2$O$_3$), a transparent conducting oxide, is a wide-band gap semiconductor with low light absorption, high electrical conductivity, and high breakdown voltage, making it a highly suitable material for advancing semiconductor technologies~\cite{zhang2022ultra, zhou2023avalanche, higashiwaki2014development, higashiwaki2012gallium,gogova2014structural, stepanov2016gallium, ahrling2019transport}.  Ga$_2$O$_3$ also serves as an excellent host for rare earth (RE) atoms with partially filled 4$f$ shell ~\cite{chen2016effects}, exhibiting various optoelectronic applications~\cite{yang2021highly, chen2020non}. RE atoms doped  oxides~\cite{grant2024Optical, thiel2011rare} exhibit unique electronic and magnetic excitations that significantly enhance quantum information processing and communication, thereby advancing quantum information science~\cite{dejneka1999rare, stevenson2022erbium, zhong2019emerging, lvovsky2009optical}. Additionally, RE doped oxide single crystals exhibit strong absorption and emission spectrum resulting in a very narrow inhomogeneous and homogeneous linewidths~\cite{lupei2008transparent, ferrier2013narrow,hughes2007lanthanide,raha2020optical, alqedra2023optical}. Particularly, erbium (Er$^{3+}$) ions in these hosts exhibit long spin and optical coherence times, and controllable 4$f$ linewidth broadening ~\cite{thiel2011rare, raha2020optical, stevenson2022erbium, grant2024Optical}, advancing both quantum memory and transduction~\cite{bhandari2023distinguishing}. The dominant physical parameter for these effects is crystalline electric field (CEF) generated from surrounding atoms, which causes to form 4$f$ multiplets of RE ions, exhibiting a sharp optical, electric and magnetic dipole transitions~\cite{liu2006spectroscopic}. 
The Er doped Ga$_2$O$_3$ emits a telecommunication wavelength of  $\sim$~1.5~$\mu$m  with life time  of 12 ms, showing a high quantum yield~\cite{vincent2008electron}, which is shorter than that in Er doped MgO but longer than those in  Er doped wider band gap hosts: SrTiO$_3$, ZnS, PbWO$_4$, TiO$_2$, ZnO, and MoO$_3$~\cite{stevenson2022erbium}. 
Further, the RE doped systems exhibit strong hyperfine interactions~\cite{rajh2022hyperfine}, which are sensitive to the CFCs~\cite{nowik1976crystal,yanovsky1975crystalline}, and are  attractive for quantum storage due to their longer hyperfine coherence time ($>$~100~$\mu$s)~\cite{goldner2006hyperfine}. The available literature primarily focuses on the photo-luminescence and optoelectronic properties of pristine and RE doped Ga$_2$O$_3$~\cite{hao2002optical, gundiah2002nanowires}, however, the fundamental understanding of phase stability, defect formation, electronic structure, and multiple
magnetic, Dzyaloshinskii-Moriya, optical, hyperfine, and crystal field interactions in Er
doped wide band gap $\alpha$- and $\beta$-Ga$_2$O$_3$ is critically important to make a foundation for both optoelectronic and quantum information applications.

\begin{figure}[!ht]
\centering	
\includegraphics[width=0.48\textwidth]{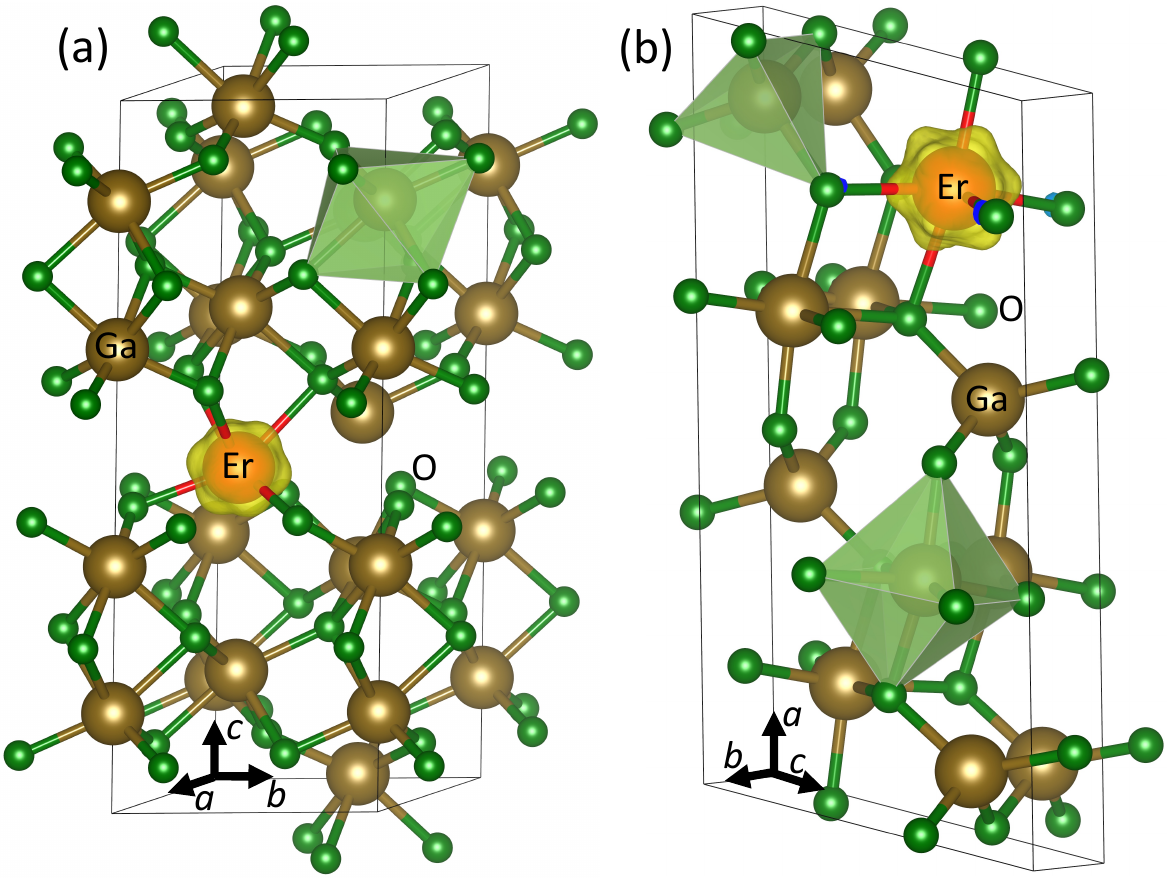}
\caption[dop]{Crystal structures of Er doped $\alpha$ (left)- and $\beta$ (right)-Ga$_2$O$_3$. In $\alpha$ phase, the Ga atoms are only octahedrally coordinated with the O atoms, while the Ga atoms are both octahedrally and tetrahedrally coordinated with the O atoms in $\beta$ phase, which are indicated by polyhedra.  In both phases, the Er atom favors  the octahedral site of Ga. The spin densities  of Er atoms are distorted by the surrounding atoms, which are presented above with isosurfaces level of 0.002~e/\AA$^3$.}
\label{figure1}
\end{figure}

The Ga$_2$O$_3$  has five distinct phases:  $\alpha$, $\beta$, $\Gamma$, $\epsilon$, and $\delta$, with  $\beta$ (monoclinic) being the most stable~\cite {roy1952polymorphism}. The $\beta$ phase  transforms to a high pressure trigonal $\alpha$ phase~\cite{MaYan-Mei_2008}. The $\beta$-Ga$_2$O$_3$ belongs to $C_\text{2h}$ point group and $C2/m$ space group  with a large  band gap of 4.6~eV to 4.9~eV depending upon the measurement method and crystal orientation~\cite{Tadjer_2019}. Both $\alpha$ and $\beta$ phases exhibit anisotropic optical properties~\cite{litimein2009fplapw}. There are two types of cation environments,  octahedral and tetrahedral,  for the Ga atom in $\beta$-Ga$_2$O$_3$~\cite{peelaers2018structural}.
The transitional metals (TMs), e.g., Mn and Ni favor octahedral site of Ga~\cite{wang2019electronic,xiao2009electronic}.  Similar to these TMs, the Er doping in $\beta$-Ga$_2$O$_3$  favors octahedral site without charge compensation defects as well as conduction electron trapping~\cite{vincent2008electron}, providing strong near infrared
emission and weak green emission peaks ~\cite{nogales2008visible}. The Er doped  Ga$_2$O$_3$ remains an insulator, however, the  band gap depends upon the Er concentration~\cite{chen2016effects}. The 4$f$ - 4$f$  transition of Er$^{3+}$ ion,  ${}^4$I$_{13/2}$ $\rightarrow$  ${}^4$I$_{15/2}$,
is 6568 cm$^{-1}$, which corresponds to the telecommunication wavelength of $\sim$~1.5~$\mu$m with life time  of 12~ms, resulting in a high quantum yield~\cite{vincent2008electron}. The peak at 550~nm (548~nm~\cite{chen2016observation}) corresponds to a green luminescence due to  ${}^4$S$_{3/2}$ $\rightarrow$ ${}^4$I$_{15/2}$ transition, which  does not shift in a wide  range of temperature from 77 to 450~K, indicating a good host material for RE~\cite{chen2016effects}.

The stable high pressure phase $\alpha$-Ga$_2$O$_3$, with $C_{3i}$ point group and $R\Bar{3}c$ space group, is an insulator with  band gap of 5.30~eV~\cite{ahmadi2019materials}. 
The Eu doped $\alpha$-Ga$_2$O$_3$ exhibits less  emission intensity than those found in $\beta$-Ga$_2$O$_3$~\cite{xie2007preparation}. The symmetric magnetic exchange interactions between  the dopants play a vital role in making a diluted magnetic semiconductor to increase the Curie  temperature ($T_C$) above room temperature~\cite{pearton2004dilute, ohno2010window}. In addition, an anisotropic exchange interaction by mirror symmetry breaking  due to spin orbit coupling (SOC), commonly known as Dzyaloshinskii-Moriya interactions (DMIs)~\cite{cahaya2022dzyaloshinskii} provide more realistic  magnetic phenomena, revealing the formation of magnetic spin spirals and skyrmions~\cite{cui2022anisotropic,caretta2020interfacial}. 
The DMI is experimentally measured in RE doped   oxides~\cite{laplane2016high}. A large orbital magnetic moment of RE and crystal field effect generated by the surrounding atoms result that the spins are not collinear, forming a canted angle between them,  exhibiting a large value of DMI of RE based systems than TM based systems~\cite{khan2019tuning,caretta2020interfacial}. Further, Er doped  Y$_2$O$_3$ exhibits a strong and anisotropic hyperfine interactions, becoming a good candidate material for the spin-based quantum applications~\cite{rajh2022hyperfine}.

In this paper, from {\textit{ab initio}}, we identify structural and chemical, mechanical, and dynamical stabilities of the pristine $\alpha$ and  $\beta$ phases of Ga$_2$O$_3$. The defect formation is also identified for Er doped Ga$_2$O$_3$ phases. Hybrid functional calculations resolve standard DFT underestimated band gaps.  The calculated  dielectric constants, refractive indexes, and absorption spectrum exhibit optical anisotropy. The magnetic exchange interactions between two Er dopants are smaller in $\beta$  than in  $\alpha$. A large value of DMI is along the $x$ direction in $\alpha$ and along the $y$ direction in $\beta$. 
Anisotropic hyperfine tensor coefficients are identified in both Er doped phases. The 4$f$ - 4$f$ transition between the lowest energy level of the first excited state to the lowest energy level of the ground state is about 1.53~$\mu$m, which falls within the quantum telecommunication wavelength regime.
	
\section{Computational and structural details}
\noindent Density functional theory (DFT) calculations, using Vienna \textit{Ab initio} Simulation Package~\cite{furthmuller1996dimer, kresse1996efficiency}, are performed to investigate phase stability, electronic and magnetic properties, and quantum phenomena of the  pristine and Er doped $\alpha$-  and $\beta$-Ga$_2$O$_3$ phases. Perdew-Burke-Ernzerhof (PBE)~\cite{perdew1996generalized} functionals of generalized gradient approximation (GGA)~\cite{gazquez2013analysis} using projector-augmented wave  (PAW) basis are used~\cite{blochl1994projector} along with a sufficient energy cutoff of 500~eV. With the PAW pseudo potential, 4$s^2$ 4$p^1$, 2$s^2$ 2$p^4$, and  4$f^{11}$ 5$s^2$ 5$p^6$ 5$d^1$ 6s$^2$ are considered as valance electrons for Ga, O, and Er atoms, respectively.  The $\alpha$ and $\beta$ phases are relaxed with optimized $k$ points grid of 4 $\times$ 4 $\times$ 1 and 2 $\times$ 8 $\times$ 4 in $\alpha$ and $\beta$, respectively, with the energy convergence threshold   of 10$^{-5}$~eV for the plane wave basis sets. 
For the density of states (DOS) calculations, dense $k$ points grids  of 4 $\times$ 16 $\times$ 8  of  $\beta$-Ga$_2$O$_3$ and 8 $\times$ 8 $\times$ 2 for $\alpha$-Ga$_2$O$_3$ are used.  The DOS are calculated using Gaussian smearing method with smearing value of 0.10~eV. The underestimated band gaps with PBE are corrected by performing hybrid functional (HSE) calculations in which we combine the exchange-correlation potential from standard DFT with the Hartree-Fock (HF) exchange. HF mixing parameter used in these calculations is 0.40~\cite{heyd2003hybrid,bhandari2023distinguishing}, which produces band gaps which are in agreement with experiments. However, for magnetic exchange interactions and DMI calculations, the strongly correlated 4$f$ electrons of Er are treated using Hubbard parameter $U_\text{eff}$(= $U$ - $J$, where $U$ = $\sim$~9.83~eV and $J$ = $\sim$~1.28~eV
are onsite Coulomb and onsite exchange potentials used in these calculations Ref~\cite{larson2007electronic}). Dynamical stability of both phases are studied from  phonon calculations using density functional perturbation theory~\cite{baroni2001phonons} with irreducible sets of 8 $\times$ 8 $\times$ 1 in $\alpha$ and 6 $\times$ 6 $\times$ 6 in $\beta$  \textit{q}-mesh, employing phonopy code~\cite{togo2013phonopy}.

As briefly mentioned in the introduction, Ga$_2$O$_3$ crystallizes with two main polymorphs: $\alpha$ (trigonal) and $\beta$ (monoclinic). The $\beta$ phase is the ground state and is lower in energy than the $\alpha$ phase by 145~meV per formula-unit. The phase transformation from $\beta$ to $\alpha$ is reported at around 19.2~GPa under cold compression. At the higher pressure, the $\alpha$ phase is experimentally confirmed to be the most stable structure~\cite{MaYan-Mei_2008}. The $\alpha$ phase consists of a hexagonal close packing of O atoms, with Ga atoms taking up two-thirds of the octahedral sites between the O atoms, and crystallizes in the trigonal $R\bar{3}c$ space group. The $\beta$ phase on the other hand, crystallizes in the monoclinic $C2/m$ space group. An equivalent Ga$^{3+}$ ions in the $\alpha$ phase are bonded to six equivalent O$^{2-}$ ions to form distorted GaO$_6$ octahedra (with three shorter (1.97~\AA) and three longer (2.09~\AA) Ga-O bonds), while two inequivalent Ga$^{3+}$ ions in the $\beta$ phase form a mixture of distorted GaO$_6$ (three shorter (1.98~\AA) and three longer (2.10~\AA) Ga-O bonds) and GaO$_4$ (1.87~\AA~ Ga-O bonds) polyhedra. A GaO$_6$ octahedra in the $\alpha$ phase shares its corners with six equivalent GaO$_6$ octahedra, while in the $\beta$ phase, a GaO$_6$ octahedra shares its corner with seven equivalent GaO$_4$ tetrahedra and edges with four equivalent GaO$_6$ octahedra~\cite{zhang2020recent}. Geometry optimization of monoclinic $\beta$- and trigonal $\alpha$-Ga$_2$O$_3$, from standard DFT calculations with PBE functionals, shows equilibrium lattice parameters, $a$ = $b$ =  5.07~\AA ~ and $c$ = 13.66~\AA~for $\alpha$-Ga$_2$O$_3$ and $a$ = 12.48~\AA, $b$ =  3.09~\AA, and $c$ =  5.89~\AA~with an angle $\beta$ =  103.68$^{\circ}$ between $a$ and $c$ for $\beta$-Ga$_2$O$_3$, which are in good agreement with available literature~\cite{shu2020electronic,dong2019elements,liu2007lattice, yoshioka2007structures,aahman1996reinvestigation,lipinska2008equation}.

For crystal field coefficients (CFCs) calculations, we  made a sufficiently large supercells of 2 $\times$ 2 $\times$ 1 in $\alpha$-Ga$_2$O$_3$ and 1 $\times$ 3 $\times$ 2 in $\beta$-Ga$_2$O$_3$. Then, a single Er atom is substituted at the favorable Ga site. The 4$f$ electrons of Er are frozen in the core, and non-magnetic (NM) calculations are performed to generate the self-consistent charge densities and local potentials~\cite{limbu2025ab, limbu2025ab1}. In this case, the 4$f$ electrons provide only a spherical contribution to the density, thereby eliminating nonspherical components arising from on-site 4$f$ states.  Consequently, the self interaction between 4$f$ electrons and their hybridization with other states are removed~\cite{novak2013crystal}, which are essential for the CFCs calculations. 

The defect formation energy of Er doped Ga$_2$O$_3$ calculated using the formula: $E_\text{f}^\text{q}(\mu_\text{i}, \epsilon_\text{F}) = E^\text{q}_\text{defect}-E_\text{pristine} + \sum_i n_i\mu_i + q(E_\text{VBM} + \epsilon_\text{F} + \Delta_\text{q/b}) + E_\text{corr}$~\cite{bhandari2023distinguishing}, where $E^\text{q}_\text{defect}$ and $E_\text{pristine}$ are the total energies of charged and pristine supercells. $E_\text{VBM}$, $\epsilon_\text{F}$, and $\Delta_\text{q/b}$ represent the valance band maxima, Fermi energy, and potential alignment between the charged and pristine supercells, respectively. This $\Delta_\text{q/b}$ is a small correction in the formation energy~\cite{bhandari2023distinguishing}. Further, $E_\text{corr}$ is a correction term, arising from the Coulomb interaction between the dopants in the supercell due to periodic boundary condition. These correction terms ($\Delta_\text{q/b}$ and $E_\text{corr}$) are calculated using SXDEFECTALIGN software~\cite{freysoldt2009fully}.
$q$ represents the charge state of the supercell, $n_\text{i}$ is a number either 1 (indicating the removal of an atom) or -1 (indicating the addition of an atom). $\mu_\text{i}$ represents the chemical potential of an atom, which depends upon the experimental growth conditions~\cite{saniz2013simplified}, and its value is bounded by the formation energy of Ga$_2$O$_3$.

\section{RESULTS AND DISCUSSION}
\subsection{Phase Stability: Structural, Mechanical and Dynamical Stability, and Defect Formation} 
The structural and chemical stabilities of $\alpha$- and $\beta$-Ga$_2$O$_3$ are investigated by employing,
$E_\text{f/c}(\text{Ga}_2\text{O}_3)$ = $E_\text{tot}(\text{Ga}_2\text{O}_3)-2E_\text{bulk/{iso}} (\text{Ga})-3{E_\text{bulk/{iso}}(\text{O})}$.
Here, $E_\text{f}$, $E_\text{c}$, $E_\text{tot}$, and $E_\text{bulk}$ and $E_\text{iso}$ represent the formation energy, cohesive energy, total energy of Ga$_2$O$_3$, and bulk and isolated energies of individual Ga and O atoms, respectively. Here the isolated energies are calculated by placing a single atom in the large cell of lattice parameter (10 \AA)\cite{limbu2022electronic}. The calculated values of formation and cohesive energies are  -1.81~eV/atom (-2.18~eV/atom~\cite{dong2019elements})  and -4.93~eV/atom for $\alpha$-Ga$_2$O$_3$, and   -1.83~eV/atom and -4.96~eV/atom (-4.33 eV/atom with hybrid functional~\cite{usseinov2021vacancy}) for $\beta$-Ga$_2$O$_3$. The negative value of cohesive and formation energies confirm the structural and chemical stabilities of  both pristine phases. 

\begin{figure}[!ht]
\centering	
\includegraphics[width=0.45\textwidth]{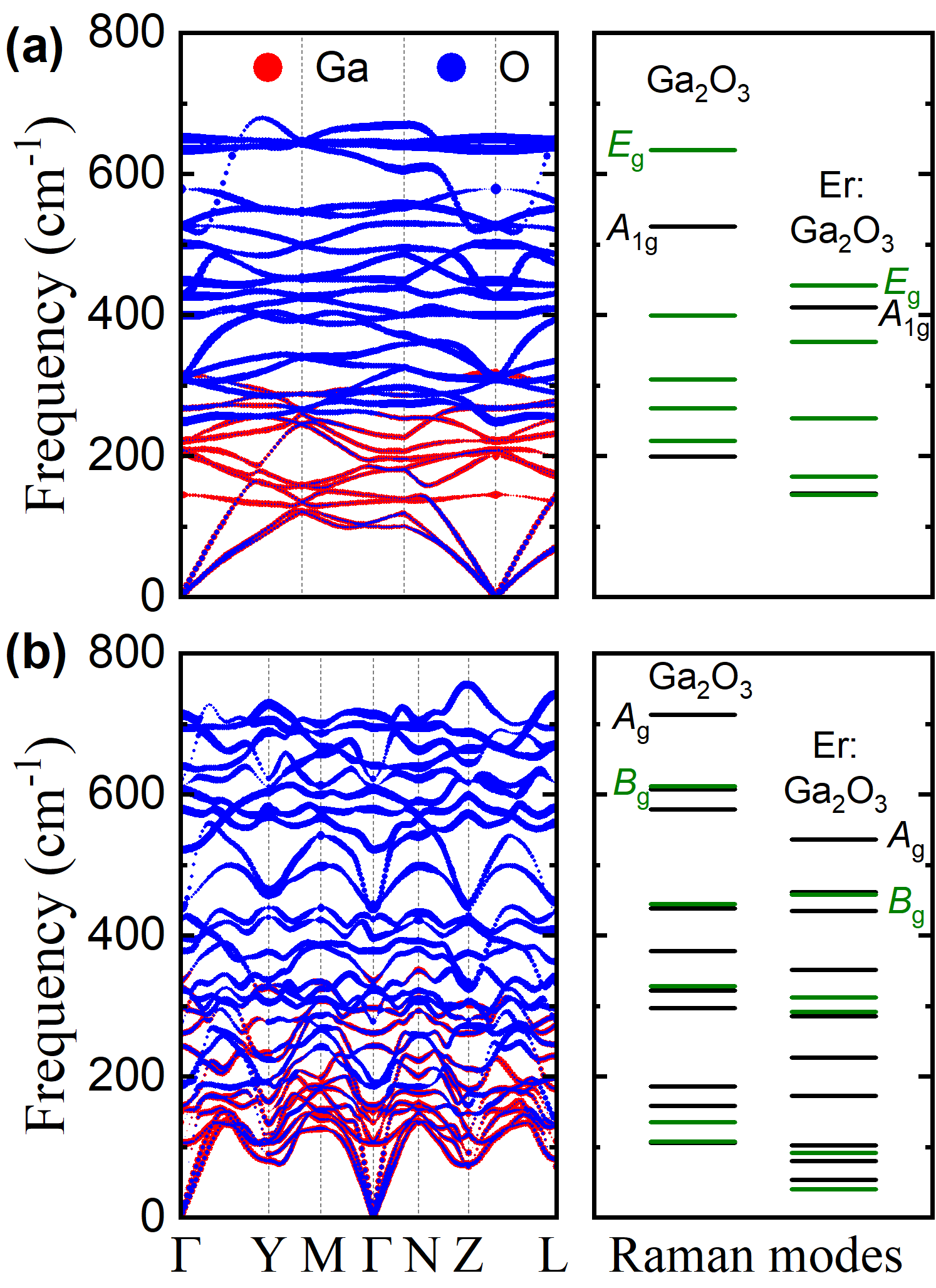}\hfill	
\caption[dop]{The phonon dispersions in pristine $\alpha$-Ga$_2$O$_3$ (a) and $\beta$-Ga$_2$O$_3$ (b). The vibrations corresponding to the Ga (O) atoms are represented in red (blue) color. The corresponding right side panels represent the frequencies of the Raman active $A_{1g}$ ($E_g$) phonon modes in the pristine and Er-doped $\alpha$-Ga$_2$O$_3$ and $A_g$ ($B_g$) modes in the pristine and Er-doped $\beta$-Ga$_2$O$_3$. The positive phonon frequencies confirm the dynamical stability.}
\label{figure2}
\end{figure}

\begin{table*}[ht!]
\centering
\caption{The GGA calculated values of elastic stiffness constants, Young's modulus (Y),  bulk modulus (B), shear modulus (G), and Poisson's ratio ($\nu$) for $\alpha$- and $\beta$-Ga$_2$O$_3$. The units of all the quantities are expressed in GPa except Poisson's ratio, which is a dimensionless quantity. In $\alpha$ phase, the elastic coefficients $C_{14}$, $C_{24}$, and $C_{56}$ satisfy the relation $C_{14}$ = - $C_{24}$ = $C_{56}$~\cite{furthmuller2016quasiparticle}}.
\setlength{\tabcolsep}{1.3 pt} 
\begin{tabular}{l c c c c c c c c c c c c c c c c c c}
\toprule
Crystals & C$_{11}$& C$_{12}$ & C$_{13}$ &  C$_{14}$ & C$_{15}$ & C$_{22}$& C$_{23}$ & C$_{25}$  & C$_{33}$ & C$_{35}$ & C$_{44}$ & C$_{46}$ &C$_{55}$&C$_{66}$ & Y & B & G & $\nu$ \\
\hline
$\alpha$-Ga$_2$O$_3$ & 384.60 & 175.10 &  136.50 & 12.18 & - & 384.60 & 136.50 & - & 341.55 & - & 89.20 & - & - & 104.75 & 262.82  &  223.00 & 100.81  & 0.30 \\
Ref.~\cite{furthmuller2016quasiparticle} & 381.5 & 173.6 & 126.0 & -17.3 & - & 381.5 & 126.0 & - & 345.8  & - & 79.7 & - & - & 104.0\\
$\beta$-Ga$_2$O$_3$ & 213.53 & 101.29 & 128.70 & - & -20.10 & 314.92  & 87.24 &  8.82 & 316.33 & 2.99 & 58.85 & 12.13 & 62.51 & 82.77 & 197.54 & 164.36 &  76.00 & 0.30 \\
Ref.~\cite{furthmuller2016quasiparticle} & 223.1 & 116.5 & 125.3 & - & -17.4 & 333.2 & 75.0 & 12.2 & 330.0 & 7.3 & 50.3 & 17.4 & 68.6 & 94.2\\
Ref.~\cite{luan2019analysis} & 206.18 & 124.08 & 123.24 & - & -2.98 & 324.18 & 63.39 & -0.41 & 297.08 & 13.89 & 39.80 & 4.67 & 83.84 & 97.23 & 187.51 & 160.13 & 71.85 & 0.30\\
\hline
\hline
\end{tabular}
\label{table1}	
\end{table*}

The calculated phonon dispersions exhibiting positive phonon frequencies confirm the dynamical stability of both $\alpha$- and $\beta$-Ga$_2$O$_3$ (Fig.~\ref{figure2}). Here, Ga atoms predominantly vibrate in the low-frequency region (below 200~cm$^{-1}$) and O atoms, which mainly vibrate at the higher frequencies, also show mixed vibrations with Ga atoms in the low-frequency region. The broad range of O atom vibrations is due to the formation of GaO$_6$ octahedra in the $\alpha$-phase and both GaO$_6$ and GaO$_4$ octahedra and tetrahedra in the $\beta$-phase. At low frequencies, these polyhedra exhibit collective motions. The Ga-O bonds are uniform in the $\alpha$-phase, while they vary in the $\beta$-phase  
due to the anisotropic polyhedral movement. The 10 atoms in the primitive cell of both $\alpha$- and $\beta$-phases result in 30 phonon branches at the center of the Brillouin zone. Among these, fifteen modes in the $\beta$-phase (10$A_g$ and 5$B_g$) and seven modes in the $\alpha$-phase (2$A_{1g}$ and 5$E_g$) are Raman active, showing distinct internal and external vibrational modes involving both Ga atoms and GaO$_x$ polyhedra in both phases. For instance, in the $\alpha$-phase, there are two $A_{1g}$ modes (at 200~cm$^{-1}$ and 527~cm$^{-1}$) primarily due to asymmetric stretching and bending along with five $E_g$ modes (at 221~cm$^{-1}$, 267~cm$^{-1}$, 309~cm$^{-1}$, 399~cm$^{-1}$, and 634~cm$^{-1}$). Similarly, $\beta$-phase shows ten $A_g$ modes (at 106~cm$^{-1}$, 158~cm$^{-1}$, 186~cm$^{-1}$, 297~cm$^{-1}$, 322~cm$^{-1}$, 378~cm$^{-1}$, 438~cm$^{-1}$, 579~cm$^{-1}$, 608~cm$^{-1}$, and 713~cm$^{-1}$) and five $B_g$ modes (at 108~cm$^{-1}$, 135~cm$^{-1}$, 338~cm$^{-1}$, 445~cm$^{-1}$, and 612~cm$^{-1}$) (Fig.~\ref{figure2}, right panel). With Er doping, one might expect the low-frequency phonon modes to soften due to the higher atomic mass of the Er atom, which is nearly three times that of the Ga atom. However, it is found that the high-frequency Raman modes soften more significantly than the low-frequency modes in both phases. For instance, in the $\alpha$-phase, the high frequency $E_g$ and $A_{1g}$ modes at 634~cm$^{-1}$ and 526~cm$^{-1}$ soften to 442~cm$^{-1}$ and 411~cm$^{-1}$, respectively. Similarly, in the $\beta$-phase, $A_g$ and $B_g$ modes at 713~cm$^{-1}$ and 612~cm$^{-1}$ drop to 536~cm$^{-1}$ and 458~cm$^{-1}$, respectively. On the other hand, the low energy $E_g$ and $A_{1g}$ modes in the $\alpha$-phase at 221~cm$^{-1}$ and 200~cm$^{-1}$ soften to 145~cm$^{-1}$ and 147~cm$^{-1}$, respectively. Further, in the $\beta$-phase, $A_g$ and $B_g$ modes at 106~cm$^{-1}$ and 108~cm$^{-1}$ drop to 53~cm$^{-1}$ and 40~cm$^{-1}$, respectively. This unusual variation in the low and high energy Raman active phonon modes is due to the existence of collective vibrations of the GaO$_x$ polyhedra rather than the individual atomic vibrations in both the $\alpha$- and $\beta$-Ga$_2$O$_3$.

Elastic stiffness coefficients are calculated in both pristine phases, applying compressive and tensile strains with the increments/decrements of 0.5\% up to 1.5\% with fully relaxed structures. These coefficients are then used to investigate mechanical stability  using mechanical stability criteria, 
C$_{11}$ $>$ $\lvert$C$_{12}$$\lvert$, C$_{44}$ $>$ 0, C$_{13}$$^2$ $<$ ${\frac{1}{2}}$C$_{33}$(C$_{11}$ + C$_{12}$), 
C$_{14}^2$ + C$_{15}^2$ $<$ $\frac{1}{2}$C$_{33}$(C$_{11}$ - C$_{12}$) = C$_{44}$C$_{66}$  for $\alpha$-Ga$_2$O$_3$, and C$_{11}$ $>$ $\lvert$C$_{12}$$\lvert$, C$_{44}$ $>$ 0, C$_{13}$$^2$ $<$ ${\frac{1}{2}}$C$_{33}$(C$_{11}$ + C$_{12}$), 
C$_{14}$$^2$  $<$ $\frac{1}{2}$C$_{33}$(C$_{11}$ - C$_{12}$) = C$_{44}$C$_{66}$ for $\beta$-Ga$_2$O$_3$ ~\cite{mouhat2014necessary}. The coefficients indeed satisfied mechanical stability criteria, confirming the mechanical stability of the pristine phases of Ga$_2$O$_3$. To further understand the mechanical properties, we calculated the Young's modulus (Y), bulk modulus (B), shear modulus (G), and Poisson's ratio ($\nu$) for both pristine phases (Table \ref{table1}). The calculated values are in good agreement with the values shown in Refs.~\cite{luan2019analysis,furthmuller2016quasiparticle} (Table \ref{table1}). The large values of Y (262.82 in $\alpha$ and 197.54 in $\beta$) indicate that these systems are not stiffer. The plastic properties of a material can be evaluated using the B/G ratio and $\nu$~\cite{luan2019analysis,pugh1954xcii}. The materials are considered ductile if B/G $>$ 1.75 and $\nu$ $>$ 0.26; otherwise, they are classified as brittle. Both phases satisfy these relations, indicating that the materials are ductile.

\begin{figure}[!ht]
\centering	
\includegraphics[width=0.48\textwidth]{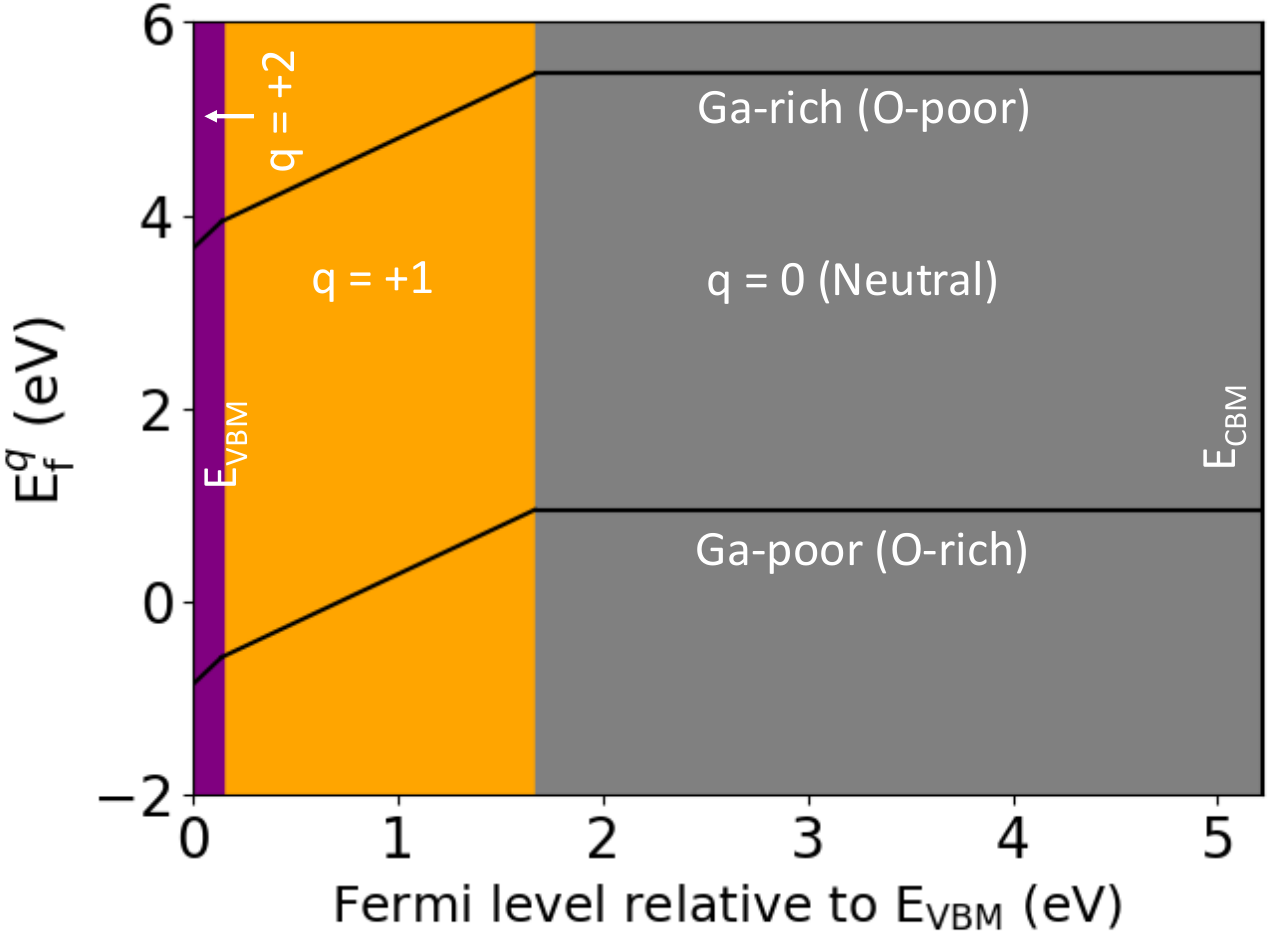}\hfill	
\caption[dop]{Defect formation energy of Er doped $\alpha$-Ga$_2$O$_3$ under two extreme growth conditions: Ga-poor (lower) and Ga-rich (upper) as a function of the Fermi level. These grow conditions are obtained by varying the chemical potential of Ga or O atoms. The vertical dotted lines indicate the different charge transitions at various Fermi levels. Further, $E_\text{VBM}$ and $E_\text{CBM}$ are the valance band maxima and conduction band minima of the pristine supercell.}
\label{formation_energy}
\end{figure}

We now analyze Er defect formation in Ga-rich (O-poor) and Ga-poor (O-rich) situations.  In extreme Ga-rich (O-poor) condition, the upper limit of the chemical potential is taken  as the bulk energy of Ga, $\mu_\text{Ga}^\text{max}$ = $\mu_\text{Ga}^\text{bulk}$, as mentioned in Ref.~\cite{saniz2013simplified}. The lower limit of the chemical potential of Ga ($\mu_\text{Ga}^\text{min}$) is calculated from the formation energy of pristine Ga$_2$O$_3$ i.e., $\mu_\text{Ga}^\text{min}$ = $\mu_\text{Ga}^\text{bulk} + E_\text{f} (\text{Ga}_2\text{O}_3)/2$. In equilibrium condition, the formation energy per formula unit and the chemical potentials of Ga and O atoms satisfy the relation,  $2\mu_{\text{Ga}}^\text{bulk}+3\mu_{\text{O}}^\text{bulk} + E_\text{f} (\text{Ga}_2\text{O}_3)$ = $2\mu_{\text{Ga}}+3\mu_{\text{O}}$. Here, we used $\mu_\text{Ga}^\text{bulk}$ = $\mu_\text{Ga}^\text{max}$ = -2.91 \text{eV}, and $\mu_\text{Ga}^\text{min}$ = -7.43 \text{eV} in $\alpha$-Ga$_2$O$_3$. Similarly,  the upper and lower limits of
 the chemical potential of O  are calculated to be $\mu_{\text{O}}^\text{bulk}$= $\mu_\text{O}^\text{max} = -4.95~ \text{eV}$ and $\mu_\text{O}^\text{min} = -7.96~ \text{eV}$. The chemical potential of Er is calculated from the cubic phase of  Er$_2$O$_3$ under extreme O-rich condition, which is found to be -16.48 \text{eV} ($\mu_\text{Er}^\text{min}$).
Then, we computed the defect formation energy of Er doped in $\alpha$-Ga$_2$O$_3$ for five different charge states: -2 (two electrons), -1 (one electron), 0 (neutral), 1 (one hole), and 2 (two holes) under two extreme experimental growth limits: Ga-rich (O-poor) and Ga-poor (O-rich). The static dielectric constant of $\alpha$ (2.74) phase is taken from Ref.~\cite{pan2022first}. In the middle of the band gap ($\epsilon_\text{F}$ = 2.61 eV), the formation energy is lower in the neutral case. 
When the Fermi level is decreased to 1.67~eV, the neutral charge state becomes a positive charge state ($q$ = +1), and it also transitions to a charge state of $q$ = +2 at the Fermi level of 0.14~eV. As expected, the formation energy is lower in the Ga-poor limit than the Ga-rich limit, indicating a higher chance of forming an Er defect under Ga-poor conditions. For the $\beta$-Ga$_2$O$_3$ phase, the  static dielectric constant (10) is taken from Ref.~\cite{fiedler2019static}. 
Similarly, the defect formation energy of Er doped in $\beta$-Ga$_2$O$_3$ is also investigated. This phase also favors a neutral charge state in the middle of the band gap. At a Fermi level of 1.90~eV, the neutral charge state transitions to a positive charge state ($q$ = +1), followed by another $q$ = +2 transition at 1.32~eV. Additionally, the neutral charge state transitions to a negative charge state ($q$ = -1) at 4.80~eV (not shown).

\subsection{ Electronic and Magnetic Properties: Band Gaps in Pristine and Magnetism in Er Doped Phases}

\begin{figure*}[!ht]
\centering	
\includegraphics[width=0.89\textwidth]{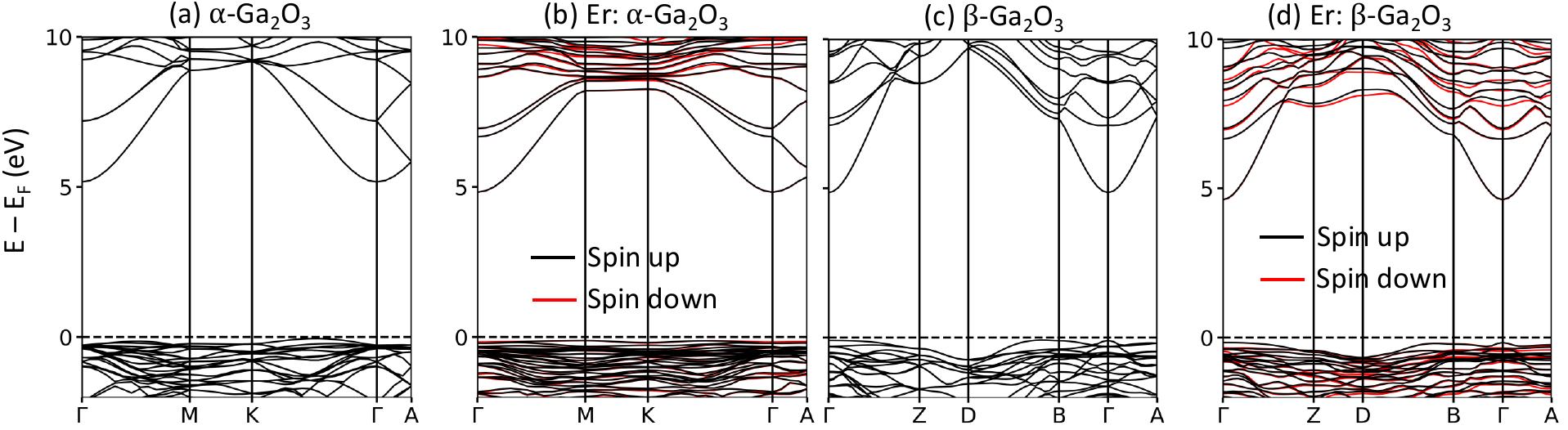}\hfill	
\caption[dop]{Band structures of pristine [(a) and (c)] and Er doped $\alpha$-Ga$_2$O$_3$ and $\beta$-Ga$_2$O$_3$ [(b) and (d)]. The Fermi level is denoted by dotted line. In both phases, we used the high symmetry points of conventional cells. In $\alpha$ phase, the high symmetry points are $\Gamma$ = (0, 0, 0),  M = (0.5, 0, 0), K = (1/3, 1/3, 0), and A = (0, 0, 0.5). Similarly, the high symmetry points used in the $\beta$ phase are  $\Gamma$ = (0, 0, 0), Z = (0, 0.5, 0),  D = (0, 	0.9927277877, 0.2536429533), B = (0, 0, 0.2536429533), and A = (-0.5200604273, 	0, 	0.1273959747).}
\label{figure3}
\end{figure*}

Standard DFT calculations show indirect band gaps of 2.33~eV and 1.98 eV in pristine $\alpha$- and $\beta$-Ga$_2$O$_3$,  which are significantly lower than the experimental values~\cite{nikolaev2017growth,ahmadi2019materials,Tadjer_2019}. The use of standard hybrid functional (HSE06) calculations correct these indirect band gaps to 4.18~eV in $\alpha$ and 3.63~eV in $\beta$-Ga$_2$O$_3$. These band gaps are also lower by $\sim$ 1.00~eV as compared to the experimental values. The hybrid functional (HSE) calculations with 40\% Hartree-Fock (HF) mixing parameter produce the band gaps of 5.21~eV in $\alpha$-Ga$_2$O$_3$ and  4.94~eV in $\beta$-Ga$_2$O$_3$ (Fig.~\ref{figure3}), which are now in good agreement with the experiment~\cite{nikolaev2017growth,ahmadi2019materials,Tadjer_2019,shinohara2008heteroepitaxy}. The region below the Fermi level, in partial density of state (PDOS), is mainly dominated from O-2$p$ states. In this region, there are  mixed states of  Ga-$s$, Ga-$p$, O-$p$, and O-$s$ with major contribution from  $p_y$ and $p_z$ states of Ga and $p_y$ and $p_z$ states of O.

\begin{figure}[!ht]
\centering	
\includegraphics[width=0.49\textwidth]{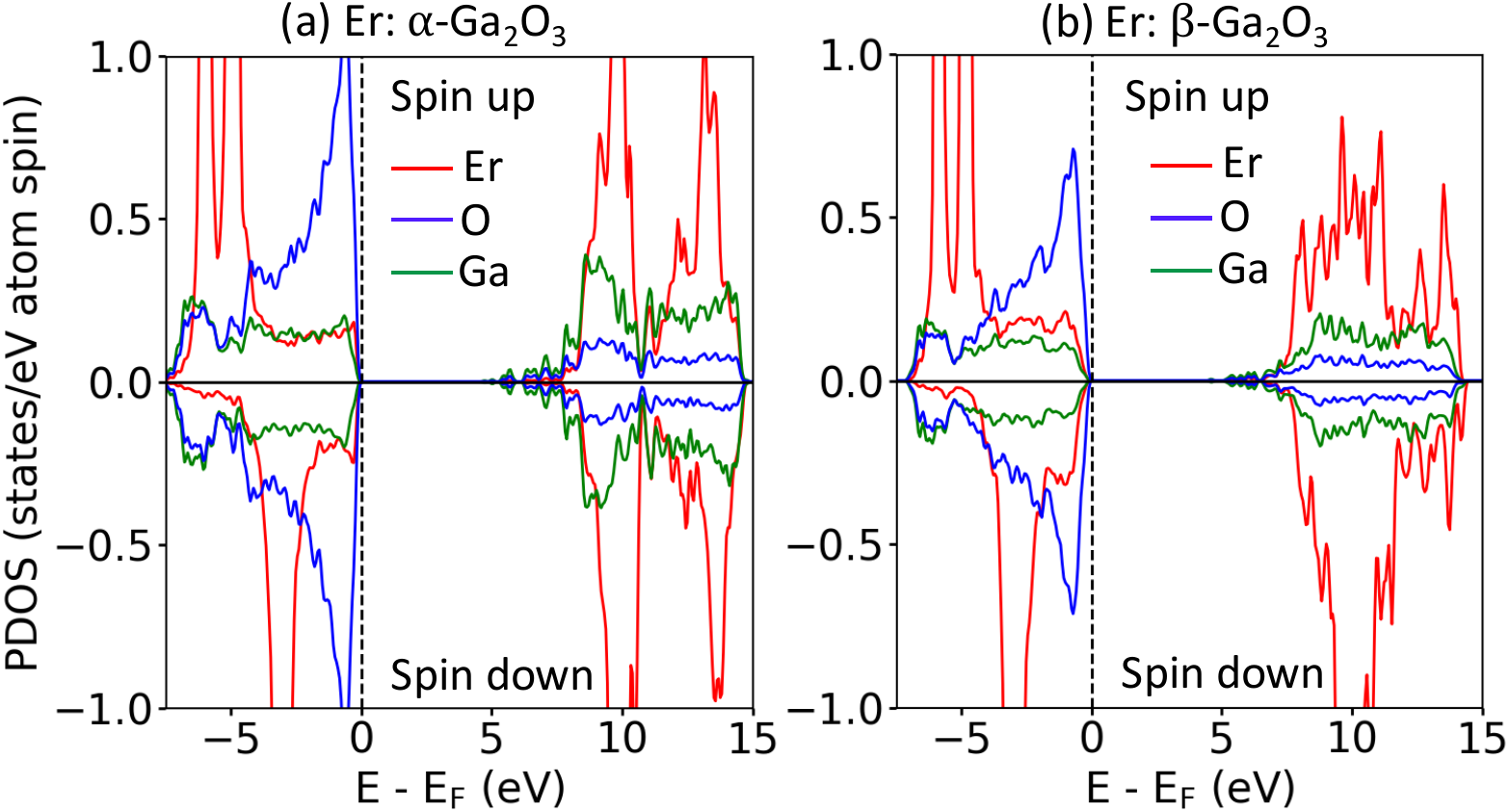}\hfill	
\caption[dop]{Partial density of states (PDOS) per atom of Er doped $\alpha$- and $\beta$-Ga$_2$O$_3$ [(a) and (b)]. The 4$f$ peaks of Er below the Fermi level are similar for both phases, which may be due to the same octahedral  environment of O atoms.}
\label{figure4}
\end{figure}

Er doped  $\alpha$-Ga$_2$O$_3$ (8.33\% doping concentration)  remains semiconductor with an indirect band gap of 4.93~eV from HSE calculations with 0.40 HF parameter, which is 0.70~eV lower than that in the pristine phase. Interestingly,  two Er doping favor antiferromagnetic (AFM) spin configuration with band gap of 5.05~eV.  Similarly, Er doped $\beta$-Ga$_2$O$_3$ (12.5\% Er doping concentration) also remains semiconductor with  band gap of  4.78~eV from HSE with 0.40 HF parameter,  which is close to the experimental band gap of 4.77~eV (7\% Er doping concentration)~\cite{chen2016effects}. We decrease the Er concentration to 6.25\% and find 4.86~eV, which is in good agreement with the experiment. Thus, the band gap of Er doped Ga$_2$O$_3$ depends on the Er doping concentration, the higher doping level shows a smaller band gap and lower doping concentration shows a larger the band gap~\cite{chen2016effects}. Moreover, two Er  doped  $\beta$-Ga$_2$O$_3$  favor ferromagnetic (FM) spin configuration with an indirect band of 4.82~eV, which is 0.04 eV higher than that found in a single  Er doped $\beta$-Ga$_2$O$_3$.

Figure~\ref{figure4} shows the PDOS of Er doped in both $\alpha$- and $\beta$ Ga$_2$O$_3$. Top of the valance band and bottom of the conduction bands are  mainly contributed from the $p$ states of O and Ga  with major contribution from the O.  The 4$f$ states of Er are in the region between -2 eV to -7 eV.  The 4$f$ DOS peak just below the Fermi level is mainly contributed from $f_{xyz}$ state, whereas the 4$f$ mainly contributed $f_{zx^2}$ is dominant above the band gap.  The 4$f$ states of Er are  strongly hybridized with $p$ states of O and Ga  with dominant contribution from O, which are also seen in charge density contour as shown in Figs.~\ref{figure1}(a) and (c). The spin magnetic moment of Er is   2.91 $\mu_B$ ($\sim$ 3 $\mu_B$) in both phases. The orbital magnetic moments of Er are 5.98 $\mu_B$ ($\sim$ 6 $\mu_B$) in $\alpha$ and 5.87 $\mu_B$ ($\sim$ 6 $\mu_B$)  in $\beta$ with GGA + U$_\text{eff}$ + SOC. Slightly smaller value of orbital magnetic moment of Er in $\mu_B$ than $\alpha$  is perhaps due to the partial quenching of orbital moment arising from low Er symmetry in $\beta$ phase. We note that the surrounding atoms of Er provide crystalline electric field, which results in the distortion of spin density and split the 4$f$ states of Er into Kramer's doublets. 

\subsection{Optical Properties}
\noindent To understand the influence of Er doping in the optical properties of Ga$_2$O$_3$, we compute the photon frequency ($\omega$) dependent dielectric functions, $\epsilon (\omega) = \epsilon_1 (\omega)+i \epsilon_2 (\omega)$, where $\epsilon_1 (\omega)$ and $\epsilon_2 (\omega)$ are real and imaginary parts of the dielectric functions ~\cite{wang2021vaspkit, gajdovs2006linear, wang2021vaspkit, gajdovs2006linear}. Refractive index is also calculated using the relation $1/{\sqrt{2}[\epsilon_1 (\omega)+\sqrt{\epsilon_1 (\omega)^2+ \epsilon_2 (\omega)^2}]^\frac{1}{2}}$~\cite{bhandari2023giant}. 
The absorption edge of polarized dipole starts at $\sim$ 5.22~eV and $\sim$ 5.25~eV along the  $z$ and $x/y$  directions in $\alpha$-Ga$_2$O$_3$, and  at $\sim$ 5.73~eV, $\sim$ 5.15~eV, and  $\sim$ 4.94~eV along the $x$, $y$, and $z$ directions, respectively,  in $\beta$-Ga$_2$O$_3$. The first prominent peak at 12.00~eV in $\alpha$-Ga$_2$O$_3$ and 13.40~eV in $\beta$-Ga$_2$O$_3$, as well as shallow peak at 19.70~eV in $\alpha$-Ga$_2$O$_3$ and 19.65~eV in $\beta$-Ga$_2$O$_3$, correspond to the inter band transition between O-2$p$ and Ga-4$s$ in first prominent peaks, and Ga-3$d$ and Ga-4$s$ in the shallow peaks, respectively. 
The real part of the dielectric functions, $\epsilon_{xx}$ = 3.07 and $\epsilon_{zz}$ = 2.97, along the $x/y$ and $z$ directions in $\alpha$-Ga$_2$O$_3$, and 3.01, 2.92, and 2.96 along the $x$, $y$, and $z$ directions  in $\beta$-Ga$_2$O$_3$ are in good agreement with Refs.~\cite{litimein2009fplapw, he2006first,passlack1995ga2o3,welch2023hybrid}. The anisotropic refractive indexes are 1.75 in the $x/y$ and 1.72 in the $z$ directions in $\alpha$-Ga$_2$O$_3$ and 1.73, 1.70, and 1.72 along the $x$, $y$, and $z$ directions in $\beta$-Ga$_2$O$_3$. Asymmetrical optical properties resulted due to the low symmetry Ga$_2$O$_3$ may support a possibility of polariton propagation~\cite{passler2022hyperbolic}.

\begin{figure}[!ht]
\centering	
\includegraphics[width=0.48\textwidth]{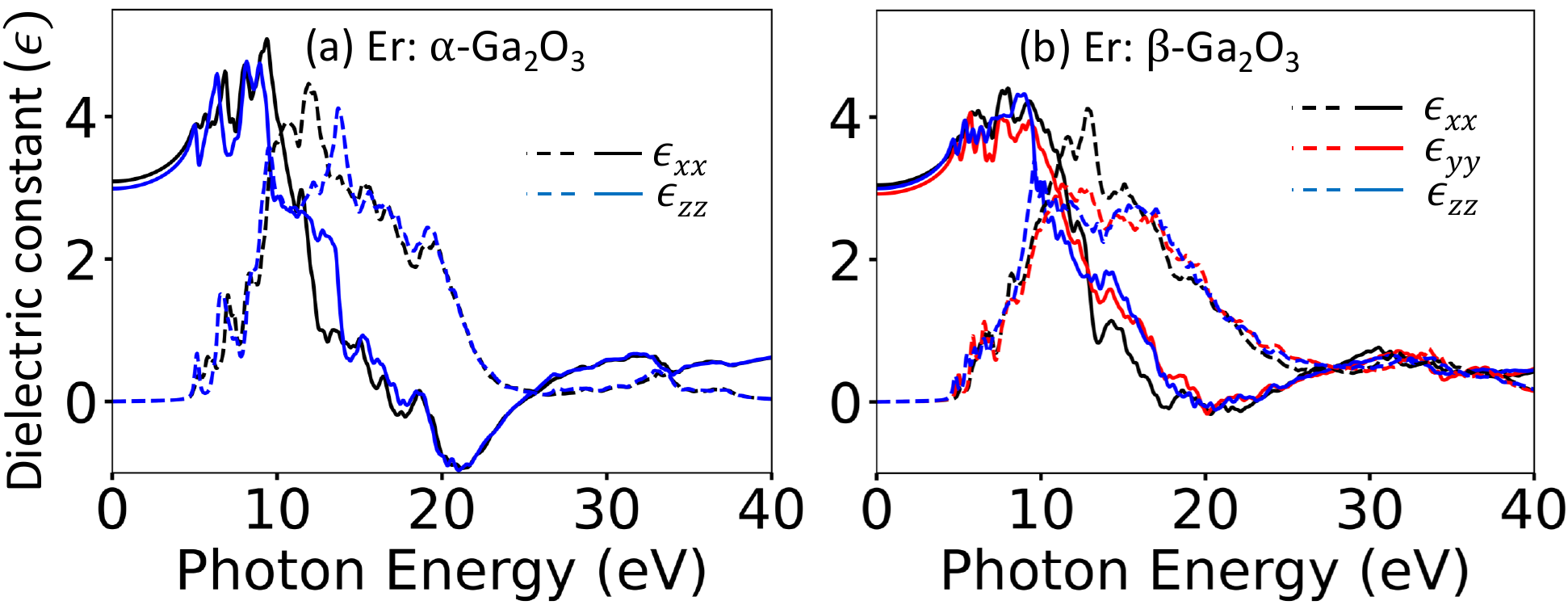}
\caption[dop]{Real (solid)  and imaginary (dotted) parts of Er doped $\alpha$-Ga$_2$O$_3$ (a) and Er doped $\beta$-Ga$_2$O$_3$.}
\label{figure5}
\end{figure}

In Er doped $\alpha$-Ga$_2$O$_3$, the absorption edges are from $\sim$~5~eV along $x$/$y$ and $z$ directions, which is slightly lower than of the the pristine phase. The absorption edges start from $\sim$ 4.63~eV along $z$, 4.87~eV along $y$, and 5.11~eV along $x$ directions in $\beta$-Ga$_2$O$_3$. In $\alpha$-Ga$_2$O$_3$, the first peak at $\sim$ 13.12~eV, from $x/y$ component of the absorption coefficient, is due to the inter-band transition between O-2$p$ and Ga-4$s$ and the second peak at 16.76~eV, $z$ component of absorption coefficient, is due to the main transition between Ga-3$d$ and Ga-4$s$ states. A small peak is obtained at $\sim$ 33.00~eV, which corresponds to the inter-band transition between Er-5$p$ and Ga-4$s$ states. In $\beta$-Ga$_2$O$_3$, the first peak at 15.79~eV is from the $x$ component of the absorption coefficient, and the second peak at 33.61~eV is from the $y$ component of the absorption coefficient. The real part of dielectric functions  are 2.47 in the $x/y$ and 2.31 in the $z$ in Er doped $\alpha$-Ga$_2$O$_3$, while in Er doped $\beta$-Ga$_2$O$_3$, the real part of dielectric functions are 2.63, 2.47, and 2.54 along the $x$, $y$, and $z$ directions. Similarly,  in Er doped $\alpha$-Ga$_2$O$_3$, the refractive indexes along the $x/y$ and $z$ directions are 1.57 and 1.52, which are smaller than those of the pristine phase. In addition, the $x$, $y$, and $z$ components of refractive indexes are 1.62, 1.57, and 1.59 in Er doped $\beta$ phase, which are smaller than those found in the pristine phase. Thus, the Er doping reduces the refractive indexes in both $\alpha$ and $\beta$ phases, potentially maximizing light propagation within the material.


\subsection{Hyperfine  Interactions}
Hyperfine interactions, due to magnetically coupled nuclear and electronic spins, produce splitting in atomic energy levels~\cite{liu2015recent} that can be calculated from DFT~\cite{yazyev2005core}. These interactions, involving a combination of Fermi-contact (isotropic) and magnetic dipole-dipole (anisotropic) interactions~\cite{defo2021calculating}, are expressed as
 ~\cite{blochl2000first}, $ A_{ij}^\text{iso}= \frac{2\mu_0\mu_e\mu_I}{3S^e S^I}\delta_{ij}\int \delta_T(r)\rho_s(r)dr$
and $A_{ij}^\text{ani}= \frac{\mu_0\mu_e\mu_I}{4\pi S^e S^I}\int \delta_T(r)\frac{\rho_s(r)}{r^3}\frac{3r_ir_j-\delta_{ij}r^2}{r^2}d^3r,$
 where {$A_{ij}^\text{iso}$} and $A_{ij}^\text{ani}$  are the isotropic and anisotropic hyperfine interaction tensors. $\mu_e$ and $\mu_I$ are the magnetic moment of nucleus with spin quantum number \textit{S$^I$} and magnetic moment of electron with spin quantum number \textit{S$^e$}. 
$\delta_T(r)$ is a smeared out
$\delta$
function. $\rho_s$(r) and  $\mu_0$ are  the charge density difference of spin up and spin down at point \textit{r} and permeability of vacuum. 
\begin{table}[!ht]
\caption{HSE calculated hyperfine interaction tensor coefficients (in~MHz) of Er doped  $\alpha$- and $\beta$-Ga$_2$O$_3$.}
\centering
\begin{tabular} {l c c c }
\hline
\hline
 & A$_{xx}$ &  A$_{yy}$ &  A$_{zz}$\\
\cline{2-4}
$\alpha$-Ga$_2$O$_3$ &  120.16 & 118.02 & 120.16 \\
$\beta$-Ga$_2$O$_3$ & 119.95 & 118.60 & 120.72\\
\hline
\hline
\end{tabular}
\label{Table3}
\end{table}

The hyperfine interaction tensor coefficients of Er doped $\alpha$- and $\beta$-Ga$_2$O$_3$ are calculated using HSE with 0.4 exact exchange mixing parameter (\ref{Table3}). In Er doped $\alpha$-Ga$_2$O$_3$, the calculated coefficients $A_{xx}$, $A_{yy}$, and $A_{zz}$ are 120.16~MHz, 118.02~MHz, and 120.16~MHz. In $\beta$-Ga$_2$O$_3$, these values are 119.95~MHz, 118.60~MHz, and 120.72~MHz. The large values of anisotropic hyperfine tensor coefficients in both cases indicate a strong interaction between the nuclear and electronic spins of Er, exhibiting electron-nuclear hybridized states in low symmetry sites~\cite{rajh2022hyperfine}.  We note that the use of bigger super cells does not change these values, indicating that the hyperfine interaction tensors are not effectively influenced by the size of the super cells. The large and strong anisotropic hyperfine interactions may make these materials excellent candidates for spin-based quantum technologies~\cite{rajh2022hyperfine}.

\subsection{Exchange and Dzyaloshinskii-Moriya Interactions (DMIs)} 
The magnetic exchange interactions of Er dopants in Ga$_2$O$_3$ are investigated by employing model Hamiltonian, 
$H_\text{spin}$ = -$J_{ab}S_a S_b$ + $D_{ab}.(S_a \times S_b)$,
where $S_a$ and $S_b$ are unit vectors, representing the local spins of $a$ and $b$ atoms and $D_{ab}$ is DMIs. The magnetic exchange interaction  ($J_{ab}$) between two Er atoms, at low doping concentration of 4.16\%, is calculated by using energy difference between anti-ferromagnetic ($E_\text{AFM}$) and ferromagnetic ($E_\text{FM}$)  configurations, $J_\text{ab}$ = $E_\text{AFM}$ - $E_\text{FM}$, as used in Refs.~\cite{mahadevan2004unusual, wierzbowska2012exchange}. The $J_{ab}$ is then calculated by placing the two Er dopants in the first nearest sites of Ga atoms. The calculated values of $J_{ab}$ is -2.02~meV in $\alpha$-Ga$_2$O$_3$  and  0.51~meV in  $\beta$-Ga$_2$O$_3$. The negative value of $J_{ab}$ in the $\alpha$ phase and the positive value in the $\beta$ phase indicate that the two Er dopants are aligned in anti-parallel spin configurations in the $\alpha$ phase and parallel spin configurations in the $\beta$ phase.

We also calculate  DMIs, $D_{ab}^z$, by considering the four different spin configurations~\cite{xiang2011predicting, yang2023first} with the $z$ direction as the spin quantization axis of the two spins at $a$ and $b$ sites: (i) $S_1$ = ($S$, 0, 0),  $S_2$ = (0, $S$, 0), (ii) $S_1$ = ($S$, 0, 0), $S_2$ = (0, -$S$, 0), (iii) $S_1$ = (-$S$, 0, 0), $S_2$ = (0, $S$, 0), and (iv) $S_1$ = (-$S$, 0, 0), $S_2$ = (0, -$S$, 0). Here, the spins of the other sites are considered the same and aligned in the $z$ direction.
{\color{black}To align the spin configurations in the desired direction, the constrained non-collinear DFT method~\cite{ma2015constrained} is employed. The total energy is then given by $E_\text{tot}$ = $E_\text{DFT}$ + $E_p$, where $E_\text{DFT}$ and $E_{p}$ correspond to DFT total energy and a penalty term. The penalty term $E_p$ enforces the alignment of local magnetic moments along the specified directions and is inversely proportional to the Lagrange multiplier $\lambda$, which must be optimized. The optimized $\lambda$ stabilizes the total energy by minimizing the $E_p$ ($\sim$~10$^{-3}$~eV in this work, consistent with  Ref.~\cite{ma2015constrained}). Thus, we first checked the total energy convergence for the desired spin configuration, selected 50 converged energies for a specific direction and computed the corresponding DMIs averaging the best-fit values to enhance accuracy. Moreover, the introduction of $E_p$ sometime alters the spin magnetic moments, requiring adjustment of the atomic sphere's radius, as mentioned in Ref~\cite{ma2015constrained}.} The spin interaction energy that includes SOC, for these four spin configurations are given by,
$E_\text{spin}$ = $D_{ab}^zS_a^{x}S_b^y$ + $E_\text{other}$. 
Then, the DMI vector along the $z$ direction is obtained from $D^z_{ab}$ = $\frac{1}{4S^2}$($E_1$ + $E_4$ - $E_2$ - $E_3$), where $E_1$, $E_2$, $E_3$, and $E_4$ are the energy corresponding to (i), (ii), (iii), and (iv) spin configurations, respectively, with the inclusion of SOC. The $x$ and $y$ components of DMI vectors are also calculated in similar manner. {\color{black}The calculated DMIs of nearest neighbor Er$^{3+}$ ions along the $x$, $y$, and $z$ components are 0.18, -0.07, and -0.11~meV 
for $\alpha$-Ga$_2$O$_3$ and  -0.09, 0.27, and -0.24~meV  
for $\beta$-Ga$_2$O$_3$}. Therefore, the highest values of DMIs are (positive in both cases) found along the $x$ direction in the $\alpha$ and along the $y$ direction in the $\beta$. 
{\color{black} The total magnitude of DMIs are 0.22~meV ($\sim$ 11\% of $J$) in $\alpha$ and 0.37~meV ($\sim$ 72\% of $J$, as reported in Ref.~\cite{laplane2016high}) in $\beta$. 
The large DMIs arise from the strong SOC of Er in both structures as compared to those of transition metal based systems where the DMIs are usually weaker than $J$~\cite{wang2023electrical,fu2024topological,camley2023consequences}}.

\subsection{Crystal Field Coefficients and Crystal Field Splitting}
The Er atom occupies Ga site and breaks the inversion symmetry $C_{3i}$ in $\alpha$-Ga$_2$O$_3$, resulting in $C_3$ local site symmetry. There are six non-zero values of $B^k_q$ coefficients: $B^2_0$, $B^4_0$, $B^4_{\pm{3}}$, $B^6_0$, $B^6_{\pm{3}}$, and $B^6_{\pm{6}}$, as also reported in Refs.~\cite{jung2019derivation, munoz1998crystal}. 
For $q$ = 3 and 6 the real part of CFCs follow the relations Re$[B^k_{{+q}}]$ = -Re$[B^k_{-q}$] and Re$[B^k_{{+q}}]$ = Re$[B^k_{{-q}}]$, respectively.   
The imaginary parts follow the relation Img$[B^k_{+q}]$ = Img$[B^k_{{-q}}]$ for $\text{q}$ = 3, and Img$[B^k_{+q}]$ = -Img$[B^k_{{-q}}]$ for $q$ = 6. The magnitude of these imaginary components $B^4_3$, $B^6_3$, and $B^6_6$ are 92.92, 1.59, and 2.59~cm$^{-1}$. The latter two values slightly change the  energy levels by $\sim$~1~cm$^{-1}$. The Er also favors octahedral site in $\beta$-Ga$_2$O$_3$, exhibiting $C_\text{s}$ local site symmetry, which exhibits in lower symmetry than $C_3$. There are fifteen non-zero values of $B_q^{k}$ for $k$ =  2, 4, and 6, and $q$ = 0, $\pm$ 1, $\pm$ 2, $\pm$ 3, $\pm$ 4, $\pm$ 5, and $\pm$ 6. These are  $B_0^{2}$, $B_{\pm{1}}^{2}$, $B_{\pm{2}}^{2}$, $B_0^{4}$, $B_{\pm{1}}^{4}$, $B_{\pm{2}}^{4}$, $B_{\pm{3}}^{4}$, $B_{\pm{4}}^{4}$, $B_0^{6}$, $B_{\pm{1}}^{6}$, $B_{\pm{2}}^{6}$, $B_{\pm{3}}^{6}$, $B_{\pm{4}}^{6}$, $B_{\pm{5}}^{6}$, and $B_{\pm{6}}^{6}$. The CFCs follow the relation Re$[B^k_{+q}]$ = Re$[B^k_{{-q}}]$ for all even $q$. But, this consistency is broken for odd $q$ (Table~\ref{Table4}). There are no imaginary components of $B_q^{k}$  in Er doped $\beta$-Ga$_2$O$_3$. Interestingly, the values corresponding to odd $q$ are not reported in Ref.~\cite{dahl1984tysonite} for $C_s$ symmetry.
\begin{table}[!ht]
\caption{Calculated crystal field coefficients (CFCs), in units of cm$^{-1}$, of Er doped $\alpha$- and $\beta$-Ga$_2$O$_3$.  The low local site symmetry of Er$^{3+}$ in $\beta$-Ga$_2$O$_3$ provides a larger number of non-zero values of CFCs.}
\centering
\begin{tabular} {l  c  c }
\hline
\hline
$B_q^k$ & $\alpha$-Ga$_2$O$_3$ & $\beta$-Ga$_2$O$_3$ \\
\hline
$B_0^2$ & -442.04 & -533.20 \\
$B_{\pm{1}}^2$ &  & $\mp$114.22\\
$B_{\pm{2}}^2$ &  & -55.77\\
$B_0^4$ & -264.10  & -269.35 \\
$B_{\pm{1}}^4$ &  & $\pm$59.86\\
$B_{\pm{2}}^4$ &  & -324.14\\
$B_{\pm{3}}^4$ & $\pm$449.90-$i$107.98  & $\mp$262.31\\
$B_{\pm{4}}^4$ &  & 217.88\\
$B_0^6$ & -3.66 & 135.94\\
$B_{\pm{1}}^6$ &  & $\pm$69.45\\
$B_{\pm{2}}^6$ &  & -44.27\\
$B_{\pm{3}}^6$ & $\pm$6.71+$i$1.84 &  $\mp$62.67\\
$B_{\pm{4}}^6$ &  & -42.23\\
$B_{\pm{5}}^6$ &  &  $\pm$52.24\\
$B_{\pm{6}}^6$ & -91.54$\mp$$i$2.63 & -156.53\\
\hline
\hline
\end{tabular}
\label{Table4}
\end{table}
\begin{figure}[!ht]
\centering	
\includegraphics[width=0.48\textwidth]{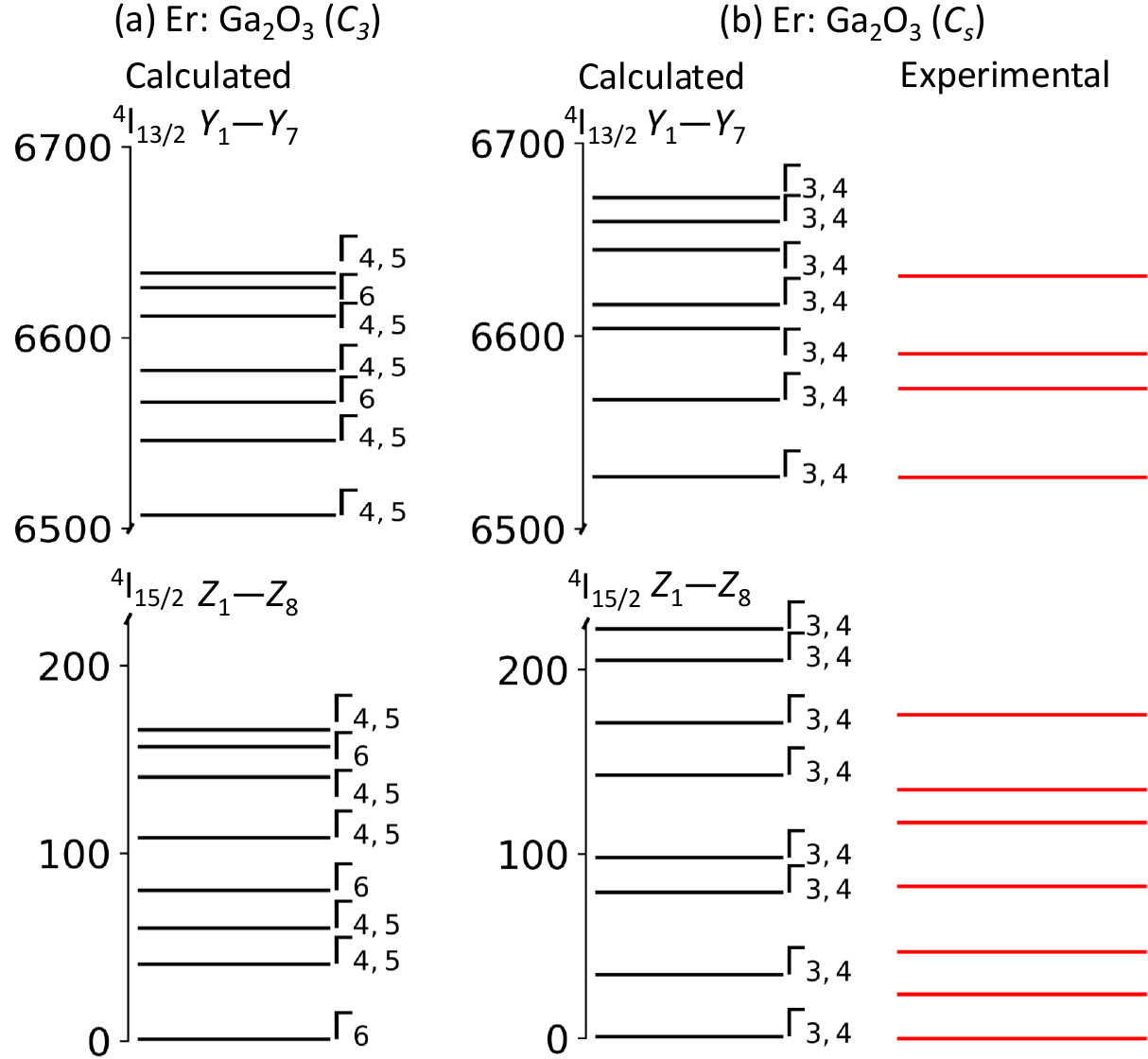}%
\caption[dop]{Crystal field  splitting of 4$f$ states of  Er$^{3+}$ in $\alpha$- and $\beta$-Ga$_2$O$_3$ [(a), (b), and (c)]. The ground and first excited states are labelled as ${}^4$I$_{15/2}$ and ${}^4$I$_{13/2}$. The calculated  first excited state energy levels are shifted by 5.88 cm$^{-1}$ (added) to match the lowest energy level of $Y_1$ (= 6527~cm$^{-1}$) with the experiment. The eight and seven doubly degenerated energy levels, due to Krammer Er$^{3+}$ ion, are found in the ground and first excited states. The ground state multiplets are labelled as $Z_i$ ($i$ = 1, 2, 3, 4, 5, 6, 7, and 8), whereas the first excited states are labelled as $Y_j$ ($j$ = 1, 2, 3, 4, 5, 6, and 7).}
\label{figure6}
\end{figure}

The local site symmetry of Er$^{3+}$ and the crystal field environment are responsible for the 4$f$ splitting of Er$^{3+}$ ion in the crystal. Without SOC and CEF effects, the Kramer ion, Er$^{3+}$, with $J$ = 15/2 has 52-fold degenerate energy levels. With SOC, these levels split into different degenerate levels: ground state (${}^4$I$_{15/2}$) and first excited (${}^4$I$_{13/2}$). Further, CEF effect splits these degenerate energy levels  into eight (5$\Gamma_{4,5}$ + 3$\Gamma_6$) and seven (5$\Gamma_{4,5}$ + 2$\Gamma_6$) multiplets in the ground and first excited state (Fig.~\ref{figure6}(a)). The combination of $\Gamma_4$ and $\Gamma_5$ form a doublet, and a single $\Gamma_6$ forms a doublet in $C_3$ symmetry. Here, we followed the irreducible representation as mentioned in Ref.~\cite{bradley1976p}. Only two irreducible representations are responsible for 4$f$ splitting of Er$^{3+}$ in the $C_3$ symmetry as also suggesed in Ref.~\cite{gruber2011crystal}. In the ground state, the multiplets $Z_1$, $Z_4$, and $Z_7$ are correspond to the $\Gamma_6$, and the remaining $Z_2$, $Z_3$, $Z_5$, $Z_6$, and $Z_8$ are correspond to $\Gamma_{4,5}$.

Similarly, in the first excited state, the multiplets $Y_1$, $Y_2$, $Y_4$, $Y_5$, and $Y_7$ correspond to $\Gamma_{4,5}$, and the multiplets, $Y_3$ and $Y_6$ correspond to $\Gamma_6$. The calculated 4$f$ - 4$f$ transitions from the lowest energy level of the first excited state to the lowest energy level of ground state, $Y_1$ $\rightarrow$ $Z_1$, is 6505.17~cm$^{-1}$ (1.53~$\mu$m), which is close to the value of 1.5~$\mu$m suggested in $\beta$-Ga$_2$O$_3$~\cite{vincent2008electron}, exhibiting $\Gamma_{4,5}$ to $\Gamma_6$ transition. To our knowledge, there are no experiments indicating energy levels of 4$f$ splitting of Er$^{3+}$  in $\alpha$-Ga$_2$O$_3$. 

Similar to $\alpha$ phase, the ground and the first excited states split into eight (8$\Gamma_{3,4}$) and (7$\Gamma_{3,4}$) seven multiplets, represented by a single irreducible representation  $\Gamma_{3,4}$ (= $\Gamma_3$ + $\Gamma_4$, where $\Gamma_3$ and $\Gamma_4$ are the complex representations) of a double group $C_s$ as mentioned in Ref.~\cite{rotereau1998vibrational} [Fig.~\ref{figure6}(b)]. The 4$f$ - 4$f$ optical transition from $Y_1$ to $Z_1$  is 6521.12~cm$^{-1}$ (1.53~$\mu$m), which is in a good agreement with the experimental value of 1.53~$\mu$m~\cite{yang2021highly, wu2016deep}.

\section{Conclusion} \noindent By performing density functional theory calculations, the phase stability, electronic and magnetic properties, and quantum phenomena of pristine and Er doped $\alpha$- and $\beta$-Ga$_2$O$_3$ phases are investigated. The negative values of cohesive and formation energies, appropriate elastic stiffness coefficients, and positive phonon frequencies confirm the structural and chemical, mechanical, and dynamical stabilities of both  pristine phases. The phonon dispersions indicate that the Ga-O bonds are uniform in the $\alpha$-phase, while they vary in the $\beta$-phase. The defect formation energy analysis confirms that both Er-doped $\alpha$- and $\beta$-Ga$_2$O$_3$ prefer Er$^{3+}$ (neutral) state. The hybrid functional calculations show an indirect band gaps of 5.21 and 4.94~eV in the pristine $\alpha$- and $\beta$-Ga$_2$O$_3$, which are in good agreement with the available experiments. A single Er substitution at the Ga site with a doping concentration of 8.33\% (one out of twelve Ga atoms) in the $\alpha$ and with a doping concentration of 12.50\% (one out of eight Ga atoms) in the $\beta$ decrease the band gaps, as compared to the pristine phases, while preserving an indirect band gap feature. Magnetic exchange interaction between two Er dopants is negative for $\alpha$ and positive for $\beta$, indicating antiferromagnetic ground state in the former and the ferromagnetic ground state in the latter. Similar values of hyperfine tensor coefficients are found for Er in both cases,  exhibiting an anisotropic nature, which could be due to the same octahedral environment.  
The large values of DMIs are found along the $x$ direction in $\alpha$ and along the $y$ direction in $\beta$. 
Calculated dielectric constant and refractive index, which are in good agreement with the experiment, exhibit an optical anisotropy. The calculated 4$f$ - 4$f$ transitions from the first excited state $Y_1$ to  the first ground state $Z_1$ are 6505.17 and 6521.12 ~cm$^{-1}$ in $\alpha$- and $\beta$-Ga$_2$O$_3$, which correspond to the telecommunication wavelength of $\sim$~1.53~$\mu$m, suggesting Er doped Ga$_2$O$_3$ is  a potential candidate material for optoelectronic and quantum telecommunication applications.

\section*{Acknowledgments}
The basic \textit{ab initio} part of the work including crystal field is supported by the U.S. Department of Energy, Office of Science, Office of Basic Energy Sciences under Award Number DE-SC0023393. The magnetic interactions and spin phonon coupling part of the work is supported as part of the Center for Energy Efficient Magnonics, an Energy Frontier Research Center funded by the U.S. Department of Energy, Office of Science, Basic Energy Sciences, under Award number DE-AC02-76SF00515. We acknowledge the use of the computational facilities on the Frontera supercomputer at the Texas Advanced Computing Center via the pathway allocation, DMR23051, and the Argon high-performance computing system at the University of Iowa.
\bibliography{references}
\end{document}